\documentclass[fleqn,twoside,twocolumn,nofootinbib,showkeys]{revtex4} 
\usepackage[sec,nocpr]{ujp} 
\begin{document}
\title[Electrical and High-Frequency Properties of Compensated
GaN]
{ELECTRICAL AND HIGH-FREQUENCY PROPERTIES\\ OF COMPENSATED GaN
UNDER ELECTRON\\ STREAMING CONDITIONS}%
\author{G.I. Syngayivska}
\affiliation{V.E.~Lashkarev Institute of Semiconductor Physics,
Department of Theoretical Physics,\\ Nat. Acad. of Sci. of Ukraine
}
\address{41, Nauky Ave., Kyiv 03028, Ukraine}
\email{koroteev@ukr.net}
\author{V.V. Korotyeyev}%
\affiliation{V.E.~Lashkarev Institute of Semiconductor Physics,
Department of Theoretical Physics,\\ Nat. Acad. of Sci. of Ukraine
}%
\address{41, Nauky Ave., Kyiv 03028, Ukraine}%
\email{koroteev@ukr.net}%
\udk{} \pacs{72.20.Ht, 72.20.Dp,\\[-3pt] 85.30.-z} \razd{\secvii}


 \autorcol{G.I.\hspace*{0.7mm}Syngaivska,
V.V.\hspace*{0.7mm}Korotyeyev }

\setcounter{page}{40}%

\begin{abstract}
Conditions required for the streaming effect and the optical-phonon
transit-time resonance to take place in a compensated bulk GaN are
analyzed in detail. Monte Carlo calculations of the high-frequency
differential electron mobility are carried out. It is shown that the
negative dynamic differential mobility can be realized in the
terahertz frequency range, at low lattice temperatures of 30--77~K,
and applied electric fields of 3--10~kV/cm. New manifestations of
the streaming effect are revealed, namely, the anisotropy of the
dynamic differential mobility and a specific behavior of the
diffusion coefficient in the direction perpendicular to the applied
electric field. The theory of terahertz radiation transmission
through the structure with an epitaxial GaN layer is developed.
Conditions for the amplification of electromagnetic waves in the
frequency range of 0.5--2~THz are obtained. The polarization
dependence of the radiation transmission coefficient through the
structure in electric fields above 1~kV/cm is found.
\end{abstract}

\maketitle

\section{Introduction}

The study of the streaming effect in semiconductors was started in
\cite{Shokl}, where it was demonstrated that a special streaming
regime of electron transport can be realized in polar semiconductors
at low temperatures and in strong enough applied electric fields.
This regime of electron transport is characterized by the emergence
of a specific quasiballistic motion of electrons in the momentum
space. In other words, under the influence of a strong electric
field, electrons are accelerated to the energy of an optical phonon,
$\hbar\omega_{0}$, almost without collisions. Having reached the
energy $\epsilon_{p}\sim\hbar\omega_{0}$, the electron stops almost
instantly and emits a polar optical phonon. Afterwards, a new cycle
of acceleration begins. Such a cyclic motion of electrons results in
the appearance of a strongly anisotropic streaming-like distribution
function of electrons in the momentum space. The distribution
function becomes strongly elongate along the electric field
direction, being almost completely confined in the passive energy
range, $\epsilon_{p}<\hbar\omega_{0}$.

For the electron streaming to be realized in semiconductor materials, a
number of conditions should be satisfied.

\noindent(I)~There must be $E^{\mathrm{(str)}}<E\ll
E^{\mathrm{(run)}}$. On the one hand, the magnitudes of applied
electric fields, $E$, must reach the values close to the
characteristic streaming field
$E^{\mathrm{(str)}}=p_{0}/e\left\langle \tau_{p}\right\rangle $,
where $p_{0}$ is the electron momentum corresponding to the optical
phonon energy, and $\left\langle \tau_{p}\right\rangle $ is the
averaged time of the electron momentum relaxation in the passive
energy range. On the other hand, the applied fields must be much
lower than a certain characteristic field of the electron runaway
effect, $E^{\mathrm{(run)}}=p_{0}/e\tau_{0}^{+}$, where
$\tau_{0}^{+}$ is the emission time of a polar optical
\mbox{phonon.}\looseness=1

\noindent(II)~For condition~I to be satisfied, the mobility of electrons in
low fields must be high enough, and, simultaneously, the interaction between
electrons and optical phonons must be strong, $\left\langle \tau
_{p}\right\rangle \gg\tau_{0}^{+}$.

\noindent(III)~The lattice temperature $T$ must be low, namely,
$k_{\rm b}T<\hbar\omega_{0}$, where $k_{\rm b}$ is the Boltzmann
constant.

\noindent(IV)~The electron concentration $n_{e}$ must be low to
avoid the electron--electron ($e-e$) scattering. At least, the time
of electron--electron collisions should  exceed $\left\langle \tau
_{p}\right\rangle $, $\tau_{e-e}>\left\langle \tau_{p}\right\rangle
$.

The theoretical calculations of electric characteristics for hot electrons under
streaming conditions meet difficulties in searching for a strongly
nonequilibrium distribution function for charge carriers. The standard
routine applied to the solution of the Boltzmann kinetic equation, which
uses the expansion of the distribution function in a series of spherical
harmonics~\cite{Dykman, Ferrytext}, is not effective, being too cumbersome,
because the harmonics of high orders (higher than the second one) have to be
taken into account. For the same reason, the conventional diffusion
approximation~\cite{Dykman} is unsuitable to describe the streaming.
The electron temperature approximation~\cite{Gantmaher} can be used only in
the case of high electron concentrations, when the $e-e$ interaction is the
dominating mechanism of scattering.

However, the approach proposed in \cite{Baraff} turned out to be rather fruitful and more adequate for the search of a strongly
anisotropic distribution function.
The Baraff method uses the approximation of a distribution
function by the sum of isotropic and needle-shaped components.
This method was widely applied in analytical researches of many types
of problems dealing with the streaming effect~\cite{Levinson, Grib, Cox}.
Note that the Baraff method, in essence, gives rise to an approximate
solution of the Boltzmann kinetic equation and provides the exact solution
only in the limit of perfect streaming, when $E\gg E^{\mathrm{(str)}}$ and
$\tau _{0}^{+}=0$.

Nowadays, the numerical methods got a wide popularity when being applied to the
solution of the Boltzmann kinetic equation. The most effective of them is the
Monte Carlo method. It enables the exact solution to be obtained for the
Boltzmann kinetic equation in a wide range of electric
fields~\cite{Boardman, Reggiani}. With the help of the Monte Carlo method, it was shown that the
streaming-like distribution function of electrons can be formed in
polar semiconductors in a dc electric field with the amplitude $E\sim
E^{\mathrm{(str)}}$~\cite{Matul, Ferry2, Peeters}.

Experimental confirmations of the formation of quasiballistic charge carrier motion in
strong electric fields were obtained in the 1980s, when studying the
current-voltage characteristics of submicronic diodes fabricated on the
basis of pure GaAs, InAs, and InP~\cite{Hickmott, Tsui}. In those
experiments, the oscillatory dependence of the static differential conductivity
on the applied electric bias was observed at low enough (helium)
temperatures, with the period of oscillations corresponding to the value of
$\hbar \omega _{0}/e$. For A$^{\mathrm{III}}$B$^{\mathrm{V}}$ compounds, the
oscillations of the static differential conductivity were observed in strong
magnetic fields as well~\cite{Hanna}.

Intensive researches of the streaming effect executed within the
last decade involved a new class of wide-bandgap semiconductor
materials, namely, group-III nitrides. Unique properties of nitride
compounds~\cite{Shur} such as, in particular, a large energy of
polar optical phonon, a large value of Fr\"{o}hlich constant, and a
relatively low effective mass (for GaN, those parameters are $\hbar
\omega _{0}/k_{\rm b}=1000$~K, $\alpha _{\rm F}=0.4$, and $m^{\ast
}/m_{e}=0.2$, respectively) considerably improve the conditions for
the streaming to take place. The Monte Carlo calculations carried
out for GaN, InN, and AlN compounds~\cite{Barry, Polyakov, Sing2}
showed that the streaming-like distribution function of electrons
emerges at temperatures of 10--150~K and in fields of 1--30~kV/cm.
It was demonstrated that the drift velocity of electrons, $V_{d}$,
and their average energy, $\langle \epsilon \rangle ,$ saturate in
this interval of fields and approach the values $V_{0}/2$
($V_{0}=p_{0}/m^{\ast }$) and $\hbar \omega _{0}/3$, respectively.

Modern researches of the streaming effect in nitrides have a pronounced
application aspect, namely, they are closely connected with the problem of
developing the terahertz radiation sources. It was shown
theoretically~\cite{Andronov1, Andronov2} that the streaming regime can be accompanied by the
emergence of a dynamic electric instability. This hypothesis was later
confirmed experimentally for InP at helium temperatures~\cite{Tulupenko}.
The frequency dependence of the dynamic (high-frequency) mobility $\mu _{\omega }
$ in a system of streaming electrons has an oscillatory alternating-sign
behavior. There exist the frequency intervals, in which $\mathrm{Re}[\mu _{\omega
}]<0$. These frequency intervals are located near the characteristic
\textit{transit-time frequency} $\nu _{R}=eE/p_{0}$. The latter corresponds to the
reciprocal electron acceleration time in the static field $E$ until the
polar optical phonon energy is reached. The appearance of the negative
dynamic differential mobility is associated with the effect of electron
bunching in the momentum space~\cite{Ishida, Shiktorov, Kozlov}.

Driven by an ac electric field with the resonance frequency $\omega \sim
2\pi \nu _{R}$, the majority of electrons move in antiphase with the
oscillations of this field, which results in the field strengthening. This
effect is called the \textit{optical phonon transit-time resonance} (OPTTR)
effect or, shortly, the \textit{transit-time resonance}. Its attractive
feature consists in that the frequency and the amplitude of this resonance can
be regulated by varying the strength of a dc electric field. This circumstance
opens wide perspectives for the creation of high-frequency sources of new
types.

Calculations of the dynamic differential mobility in doped bulk
nitrides~\cite{Varani1,Varani2} and high-quality nitride quantum
wells~\cite{Varani3, JCAO} showed that the dynamic negative
differential mobility (DNDM) can reach several hundreds of
$\mathrm{cm}^{2}$/V/s in the frequency range from 0.5 to a few
terahertz, in electric fields of 1--10~kV/cm, and within the
temperature interval of 10--77~K. Similar conditions for the DNDM in
GaN quantum wells were obtained in works~\cite{Korotyeyev1,
Korotyeyev2}. Note, however, that the $e-e$ scattering was not taken
into account in those calculations. The electron-electron coupling
can change the conditions of existence for the DNDM very much. For a
GaN quantum well, it was shown~\cite{JCAOee} that the DNDM amplitude
substantially decreases already at the electron concentration
$n_{e}=10^{11}$~\textrm{cm}$^{-3}$ (the extrapolation of this value
to the bulk sample gives the critical value of electron
concentration $n_{e}=10^{16}\div 10^{17}$~\textrm{cm}$^{-3}$). At
high electron concentrations, when the $e-e$ scattering dominates,
the DNDM does not appear~\cite{Sokolov}. The negative influence of
the electron-electron scattering can be avoided with the help of the
compensation of free carriers. A high compensation degree allows the
electron concentration to be reduced, hence making the
electron-electron scattering not substantial. The case of
compensated GaN was not discussed in detail in the literature.
However, this case is important, because it provides better
conditions for the streaming and transit-time resonance effects to
be observed.

The main purpose of our researches was to reveal additional features in the
electric characteristics of compensated GaN, which could definitely testify
to the streaming emergence and could be identified in future experiments. In
particular, we calculated, for the first time, the field dependence of
the transverse diffusion coefficient and the frequency dependences of nonzero
components of the dynamic mobility tensor in strong enough electric fields. It
is of importance that those dependences can be observed in electro-gradient
experiments and optical experiments dealing with the transmission of
electromagnetic radiation with a given polarization through a structure that
contains a layer of compensated GaN. For the calculations of stationary and
high-frequency parameters of the electron gas in strong electric fields, the
Monte Carlo numerical method was applied.

The paper is organized as follows. In Section~\ref{secII}, the model
of electron transport is described. In Section~\ref{secIII}, the
features in the electron distribution function that emerges in a
constant electric field are discussed, and the dependences of
electric parameters of the electron gas on the amplitude of an
applied field are analyzed. In Section~\ref{secIV}, the effect of
transit-time resonance is studied, the spectra of the high-frequency
mobility are presented for various relative orientations of dc and
high-frequency electric fields, and the existence conditions for the
DNDM are analyzed. In Section~\ref{secV}, the theory of terahertz
radiation transmission through a structure with a thin epitaxial GaN
layer is developed. The main conclusions are summarized in
Section~\ref{secVI}.

\section{Model of Electron Transport}

\label{secII} Bulk GaN with a cubic modification and with the
concentration of ionized impurities
$N_{i}=10^{16}$~\textrm{cm}$^{-3}$ is considered. The concentration
of electrons is supposed to be $n_{e}<N_{i}$, i.e. the semiconductor
is suggested to be compensated. The electron transport was simulated
using the single-particle Monte Carlo method. The basic algorithms
applied at the simulation were standard; they are described in
detail in works~\cite{Reggiani, Boardman}. In our model, we
considered the processes of electron scattering by acoustic and
polar optical phonons, as well as by ionized impurities. The
dispersion law for electrons was assumed to be parabolic, and all
the processes of electron scattering were supposed to occur only
near the bottom of the lowest $\Gamma $-valley. Explicit expressions
for the probabilities of electron scattering by acoustic and polar
optical phonons can be found in works~\cite{Reggiani, Kochelap}. The
electron scattering by ionized impurities was considered within the
approach described in work~\cite{Reggiani}; it was found to be more
correct for compensated semiconductors in comparison with the
conventional Brooks--Herring and Conwell--Weisskopf models. This
approach is described in Section~\ref{sec2.1} in more details.

\subsection{Electron scattering by ionized impurities}

\label{sec2.1}As a rule, the electron scattering by ionized impurities is
considered in the framework of either the Brooks--Herring (BH) or the
Conwell--Weisskopf (CW) model. In these models, the screening of the field,
which is induced by impurity ions, by conduction electrons is taken into
account in different ways. In particular, the BH model uses the screened Coulomb
potential
\begin{equation}
V(r)=\frac{Z_{i}e}{\kappa _{0}r}\exp(-r/\lambda _{\rm D}),
\label{Kulon}
\end{equation}%
where $Z_{i}e$ is the charge of an impurity ion, $\kappa _{0}$ is the
dielectric permittivity, $\lambda _{\rm D}=(\kappa _{0}k_{\rm
b}T/4\pi e^{2}n_{e})^{1/2}$ is the Debye screening length, and
$n_{e}$ is the electron concentration. In the CW model, the
unscreened Coulomb potential is cut off at the distance $b=(3/4\pi
N_{i})^{1/3}$ between impurity ions, and the minimum scattering
angle $\theta _{\mathrm{min}}$ for an electron with energy
$\epsilon $ is determined by the formula
\begin{equation}
{\rm ctg}\left(\! \frac{\theta _{\mathrm{min}}}{2}\!\right)
=\frac{2\epsilon b\kappa _{0}}{e^{2}}.
\end{equation}%
Which of those models should be used depends on the ratio between
$\lambda _{\rm D}$ and $b$. In a semiconductor with a high
compensation degree ($n_{e}\ll N_{i}$), the inequality $\lambda
_{\rm D}\gg b$ is obeyed so that the CW model proves to be more
adequate for applications. For a heavily doped semiconductor, in
which all impurities are ionized ($N_{i}=n_{e}$), the inverse
inequality $\lambda _{\rm D}\ll b$ can be valid. In this case, the
BH model is reasonable to be~used.

In the cases where $\lambda _{\rm D}\sim b$, it was
suggested \cite{Reggiani} to use the CW model with the screened Coulomb
potential, rather than the \textquotedblleft pure\textquotedblright\
one. In this model, the probability for an electron to
transit from the initial state described by the wave vector ${\bf k}$
into the state with the wave vector ${\bf k}^{\prime }$ within a unit
time interval, provided that the electron is scattered by ionized
impurities, is given by the expression
\[
W_{{\bf k},{\bf k}^{^{\prime }}}=\frac{2^{5}\pi
^{3}e^{4}Z_{i}^{2}N_{i}}{\hbar \kappa _{0}^{2}\Omega }\left(\!
\lambda _{\rm D}^{-2}+({\bf k}^{^{\prime }}-{\bf k})^{2}\!\right)
^{-2}\times
\]\vspace*{-5mm}
\begin{equation}
\times\delta (\epsilon _{{\bf k}^{^{\prime }}}-\epsilon _{{\bf k}}),
\label{prob}
\end{equation}%
where $\Omega $ is the normalization volume. In order to calculate
the probability of the electron transition from the initial state
into any other one within a unit time interval (the scattering
rate), the quantity $W_{{\bf k},{\bf k}^{\prime }}$ has to be
multiplied by $\Omega /(2\pi )^{3},$ and the product has to be
integrated over all ${\bf k}^{\prime }$-values, bearing in mind that
the angle $\theta $ between the vectors ${\bf k}$ and ${\bf
k}^{\prime }$ changes from $\theta _{\mathrm{min}}$ to $\pi $. For
the parabolic dispersion law, we obtain the following formula
describing the scattering rate for an electron with
\mbox{energy~$\epsilon $:}\vspace{-3mm}
\begin{equation}
r_{\mathrm{imp}}=\frac{2^{1/2}\pi Z_{i}^{2}e^{4}N_{i}}{\kappa
_{0}^{2}m^{\ast 1/2}\epsilon ^{1/2}}\left[ \frac{1}{\epsilon _{\rm
D}+4\epsilon \sin ^{2}\frac{\theta
_{\mathrm{min}}}{2}}-\frac{1}{\epsilon _{\rm D}+4\epsilon
}\right]\!, \label{inter}
\end{equation}\vspace{-5mm}%

\noindent where $\epsilon _{\rm D}=\hbar ^{2}/2m^{\ast }\lambda
_{\rm D}^{2}$. Formula (\ref{inter}) gives the limiting transitions
to the BH model as $\theta _{\mathrm{min}}\rightarrow 0$ and to the
CW one as $\epsilon _{\rm D}\rightarrow 0$.

In a compensated GaN with the concentration of ionized impurities
$N_{i}=10^{16}$~\textrm{cm}$^{-3}$, the concentration of electrons
$n_{e}=10^{15}$~\textrm{cm}$^{-3}$, and at a temperature of 30~K
-- below, it will be demonstrated that these parameters are the
best for the realization of the OPTTR -- the values
$b=28$\textrm{~nm }and $\lambda _{\rm D}=35$\textrm{~nm} turn out
close to each other so that it is formula (\ref{inter}) that
should be used.

\subsection{\vspace*{-0.5mm}Total scattering rate}

In our transport model, the total scattering rate is equal to the
sum
$r_{\mathrm{tot}}=r_{\mathrm{ac}}+r_{\mathrm{imp}}+r_{\mathrm{op}}$,
where $r_{\mathrm{imp}}$, $r_{\mathrm{ac}}$, and $r_{\mathrm{op}}$
are the probabilities of the electron scattering by ionized
impurities, acoustic phonons, and polar optical phonons,
respectively. The probabilities
$r_{\mathrm{ac}}=r_{\mathrm{ac}}^{+}+r_{\mathrm{ac}}^{-}$ and
$r_{\mathrm{op}}=r_{\mathrm{op}}^{+}+r_{\mathrm{op}}^{-}$ take into
account the emission ($+ $) and absorption ($-$) processes of
acoustic and polar optical phonons, respectively. Note that, within
the actual temperature interval, the electron scattering by acoustic
phonons is almost elastic so that $r_{\mathrm{ac}}^{+}\sim
r_{\mathrm{ac}}^{-}$. At the same time, the mechanism of electron
scattering by polar optical phonons is essentially inelastic.
Therefore, the relation between the probabilities
$r_{\mathrm{op}}^{+}$ and $r_{\mathrm{op}}^{-}$ strongly depends on
the lattice temperature and the electron~energy.

Figure~1 demonstrates the dependences of the total probability of
electron scattering, $r_{\mathrm{tot}}$, on the electron energy in
bulk GaN with the concentration of ionized impurities
$N_{i}=10^{16}$~\textrm{cm}$^{-3}$ and the concentration of
electrons $n_{e}=10^{15}$~\textrm{cm}$^{-3}$ calculated for two
lattice temperatures, $T=30$ and 300~K. One can easily see a large
difference between $r_{\mathrm{tot}}$-values in the passive
($\epsilon <\hbar \omega _{0}$) and active ($\epsilon >\hbar \omega
_{0}$) energy regions. It can be explained by the fact that
the electrons with the energy $\epsilon <\hbar \omega _{0}$ are mostly
scattered by ionized impurities and acoustic phonons, whereas the
main scattering process for electrons with $\epsilon >\hbar \omega
_{0}$ is driven by the more intensive spontaneous emission of polar
optical phonons. For instance, for thermal electrons, i.e. electrons
with the energy $\epsilon =0.028\times \hbar \omega _{0}$, the
probabilities of their scattering by ionized impurities and acoustic
phonons equal $r_{\mathrm{imp}}=1.6\times 10^{12}$~s$^{-1}$ and
$r_{\mathrm{ac}}=1.6\times 10^{19}$~s$^{-1}$, respectively, at a
lattice temperature of 30~K (curve~{\it 1} in Fig.~1). For an
electron with the energy $\epsilon =1.2\times \hbar \omega _{0}$,
the probability of the polar optical phonon emission is
$r_{\mathrm{op}}^{+}=4.5\times 10^{13}~\mathrm{s}^{-1}$, whereas
$r_{\mathrm{ac}}=10^{11}$~s$^{-1}$ and $r_{\mathrm{imp}}=7.5\times
10^{12}$~s$^{-1}$. At the same time, the polar optical phonon
absorption is practically absent, $r_{\mathrm{op}}^{-}\sim
10^{2}$~s$^{-1}$.

\begin{figure}
\includegraphics[width=\column]{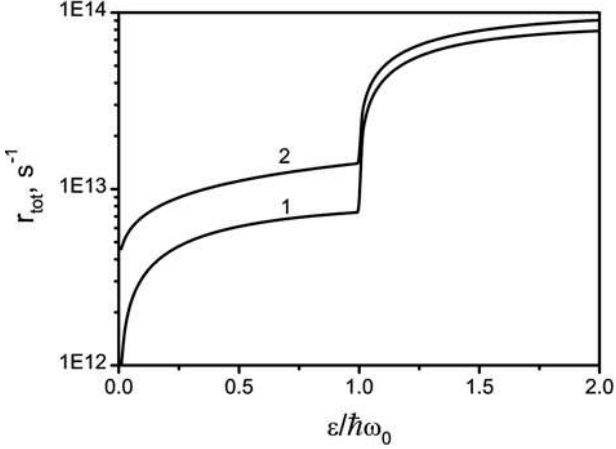}
\vskip-3mm\caption{Dependences of the total electron scattering rate
on~the electron energy at the temperature $T=30$ ({\it 1}) and
300~K~({\it 2})  }
\end{figure}

At room temperature of the lattice, $T=300$~K (curve~{\it 2} in
Fig.~1), the value of $r_{\mathrm{tot}}$ for electrons with the
energy $\epsilon <\hbar \omega _{0}$ turns out several times larger
owing mainly to the growth of the role of inelastic processes in the
optical phonon absorption. For example, for electrons with the
energy $\epsilon =0.28\times \hbar \omega _{0}$, we have
$r_{\mathrm{imp}}=5.5\times 10^{12}$~s$^{-1}$,
$r_{\mathrm{ac}}=5.1\times 10^{11}$~s$^{-1}$, and
$r_{\mathrm{op}}^{-}= 3.3\times 10^{12}$~s$^{-1}$. At the same time,
for an electron with the energy in the active range,
$\epsilon=$\linebreak $ =1.2\times \hbar \omega _{0}$, we have
$r_{\mathrm{ac}}=10^{12}$~s$^{-1}$,
$r_{\mathrm{imp}}=10^{13}$~s$^{-1}$, $r_{\mathrm{op}}^{-}=3\times
10^{12}$~s$^{-1}$, and $r_{\mathrm{op}}^{+}=4.5\times
10^{13}$~s$^{-1}$.

The dependence of $r_{\mathrm{tot}}$ on the electron energy, which
was described above, is inherent to all nitrides. Note that the
larger the difference between the $r_{\mathrm{tot}}$-values in the
active and passive energy ranges, the more favorable are the
conditions for the emergence of the streaming effect.

\section{\vspace*{-0.5mm}Steady-State Electron Characteristics}

\label{secIII}In the streaming regime, the electron motion becomes
quasiperiodic, which finds its reflection in the distribution function of
electrons and in the transport characteristics of the electron gas. The
qualitative estimation of the characteristic electric field
$E^{\mathrm{(str)}}$, in which the streaming regime is realized, can be done on the basis of
low-field mobility values for electrons. Therefore, instead of immediately
proceeding to the analysis of stationary electron parameters, it is
reasonable to discuss the dependence of the low-field electron
mobility on the lattice temperature in detail.

\subsection{\vspace*{-0.5mm}Low-field electron mobility}

At low temperatures ($k_{\rm b}T<\hbar \omega _{0}$), when only the
elastic mechanisms of scattering are actual, the electron mobility
can be calculated analytically with the use of the $\tau $-approximation,
namely,
\begin{equation}
\mu _{0}=\frac{e}{m^{\ast }}\langle \tau _{p}\rangle,  \label{mob}
\end{equation}
where
\begin{equation}
\langle \tau _{p}\rangle =\frac{4}{3\sqrt{\pi }(k_{\rm
B}T)^{5/2}}\!\int\limits_{0}^{\infty }\!\!d\epsilon \epsilon
^{3/2}r_{p}^{-1}(\epsilon )\exp \left(\! -\frac{\epsilon }{k_{\rm
B}T}\!\right)   \label{tau_p}
\end{equation}

\noindent is the statistically averaged   momentum relaxation time.
Here, $r_{p}$ is the inverse time of momentum relaxation. According
to Matthiessen's rule,
\begin{equation}
r_{p}=r_{p,\mathrm{ac}}+r_{p,\mathrm{imp}},
\end{equation}%
where $r_{p,\mathrm{ac}}$ and $r_{p,\mathrm{imp}}$ are the inverse
 momentum relaxation time at acoustic phonons and ionized
impurities, respectively. The explicit expressions for
$r_{p,\mathrm{ac}}$ and $r_{p,\mathrm{imp}}$ can be found in many
manuals (see, e.g., textbook~\cite{Bonch}). If the process of
scattering by ionized impurities is examined in the framework of the
Conwell--Weisskopf model with screened Coulomb potential, the
following formula has to be used for the calculation of
$r_{p,\mathrm{imp}}$:
\[
r_{p,{\rm imp}}=\frac{\pi
Z_{i}^{2}e^{4}N_{i}}{\kappa_{0}^2\sqrt{2m^{*}}}
\frac{1}{\epsilon^{3/2}} \left[\log\left(\!\frac{\epsilon_{\rm
D}+4\epsilon} {\epsilon_{\rm
D}+4\epsilon\sin^{2}\frac{\theta_{\min}}{2}}\!\right)- \right.
\]\vspace*{-5mm}
\begin{equation}
\left. \frac{4\epsilon\epsilon_{\rm D}
(1-\sin^{2}\frac{\theta_{\min}}{2})} {(\epsilon_{\rm
D}+4\epsilon)(\epsilon_{\rm
D}+4\epsilon\sin^{2}\frac{\theta_{\min}}{2})}\right]\!.
\label{rp_imp}
\end{equation}

The dependence of $r_{p,\mathrm{imp}}$ on the electron energy $\epsilon $ is
shown in the inset in Fig.~2. At high energies, the value of
$r_{p,\mathrm{imp}}$ decreases with the growth of $\epsilon $ as $\epsilon ^{-3/2}$. Such
a behavior of $r_{p,\mathrm{imp}}$ is explained by the fact that faster
electrons are mainly scattered at small angles. For the sake of comparison,
the probability of the electron scattering by ionized impurities
$r_{\mathrm{imp}}$ grows with the electron energy (see Fig.~1).

Note that, at high temperatures, when the role of the inelastic
scattering mechanisms becomes essential, the $\tau $-approximation
loses its meaning, so that expression (\ref{mob}) cannot be used. For
the exact calculation of the electron mobility in a low field, $\mu
_{0}$, in a wide temperature range, either of the following
procedures can be applied: (1)~the Monte Carlo method is used to
calculate the dependence of electron drift velocity
$V_{\mathrm{dr}}$ in the field with strength $E$; afterwards,
the electron mobility can be determined from the slope of the curve
$V_{\mathrm{dr}}(E)$~\cite{Gupta}; or (2)~the Monte Carlo method is
used to calculate the diffusion coefficient $D_{0}$; afterwards, the
electron mobility can be determined from the Einstein relation $\mu
_{0}=eD_{0}/k_{\rm b}T$~\cite{Littlejohn}. In the case of low
fields, the second way turns out to be more accurate and less
dependent on the statistical noise produced by the Monte Carlo
calculations. Therefore, it was the second way that was selected by
us to calculate the dependence of the electron mobility in low fields on
the lattice temperature.

As is seen from Fig.~2, the dependence of the low-field mobility $\mu _{0}$ on $T
$ is nonmonotonous. In the temperature interval 30--120~K, the $\mu
_{0}$-values increase with $T$, because electrons are mainly scattered by
acoustic phonons and ionized impurities, with the latter process dominating.
It is of interest that the growth of $\mu _{0}$ at higher $T$ is associated
with the decreasing dependence of $r_{p,\mathrm{imp}}$ on the energy. As the
lattice temperature increases, the fraction of high-energy electrons, for
which scattering by ionized impurities is less intensive, grows. In Fig.~2,
the results of calculations of the low-field mobility carried out either in the
framework of the Monte Carlo method or with the use of formulas
(\ref{mob})--(\ref{rp_imp}) are evidently identical within the temperature interval
$T=30-120$~K. Starting from a temperature of 120~K, the electron mobility $\mu _{0}$
diminishes with the growth of $T$, which is connected with the enhancement
of the role of the mechanisms of scattering by acoustic and polar optical
phonons. If the temperature continues to grow, the role of the electron
scattering by polar optical phonons increases considerably. In Fig.~2, this
fact is illustrated by an increasing discrepancy between the exact mobility
value obtained by the Monte Carlo method and its approximation calculated by
formulas (\ref{mob})--(\ref{rp_imp}). At room temperature, the mobility of
electrons is mainly governed by their scattering by polar optical phonons.
The values calculated by us are close to those measured in high-quality GaN
epitaxial layers grown up on Al$_{2}$O$_{3}$ substrates~\cite{Look}.

At $T=30$~K, the electron mobility equals
5000~$\mathrm{cm}^{\mathrm{2}}$/V/s so that the characteristic electric field of the streaming is
$E^{\mathrm{str}}=8$~kV/cm; whereas the electron mobility at $T=77$~K is close to
10000~$\mathrm{cm}^{\mathrm{2}}$/V/s, and $E^{\mathrm{str}}=4$~kV/cm.

\begin{figure}
\includegraphics[width=\column]{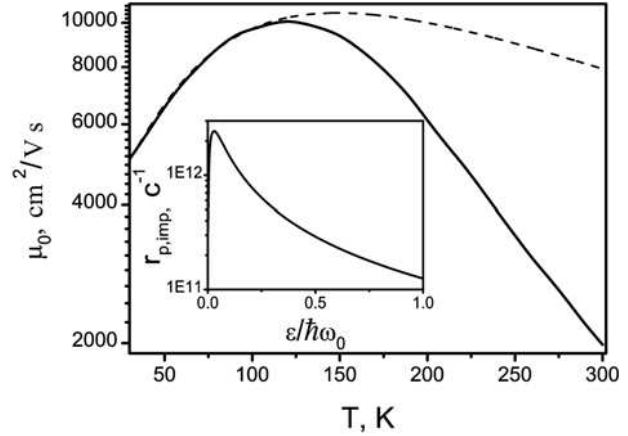}
\vskip-3mm\caption{Dependences of the low-field electron mobility on
the lattice temperature calculated exactly by the Monte Carlo method
(solid curve) and approximately by formulas
(\ref{mob})--(\ref{rp_imp}) (dashed curve).
$N_{i}=10^{16}$~\textrm{cm}$^{-3}$ and
$n_{e}=10^{15}~\mathrm{cm}^{-3}$  }
\end{figure}

\subsection{Electron distribution function}

Generally speaking, the distribution function of electrons in a bulk
semiconductor is a function of three momentum variables,
$F(p_{x},p_{y},p_{z})$. However, the distribution function of electrons in a
uniform dc electric field has the axial symmetry with respect to the field
direction. Hence, if an electric field directed along the $z$-axis is
applied to the semiconductor, it is enough to analyze the distribution
functions of electrons in the momentum space in two directions, namely,
along the field, $f(p_{z})$, and across it, $f(p_{x})$. The distribution
functions $f(p_{z})$ and $f(p_{x})$ are introduced as follows:
\begin{equation*}
f(p_{z})=\int \int dp_{x}dp_{y}F(p_{x},p_{y},p_{z})/N
\end{equation*}and \begin{equation*}
f(p_{x})=\int \int dp_{z}dp_{y}F(p_{x},p_{y},p_{z})/N,
\end{equation*}where \begin{equation*}
N=\int \int \int dp_{x}dp_{y}dp_{z}F(p_{x},p_{y},p_{z})
\end{equation*}
is the normalization integral.

\begin{figure}
\includegraphics[width=\column]{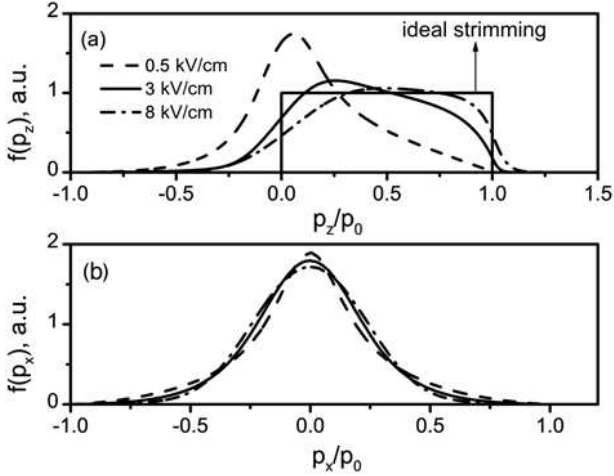}
\vskip-3mm\caption{Distributions of electrons in the momentum space
({\it a})~along and ({\it b})~across the electric field in GaN.
$N_{i}=$ =~$10^{16}$~\textrm{cm}$^{-3}$,
$n_{e}=10^{15}~\mathrm{cm}^{-3}$, and $T=30~\mathrm{K}$  }\vskip3mm
\end{figure}

\begin{figure}
\includegraphics[width=\column]{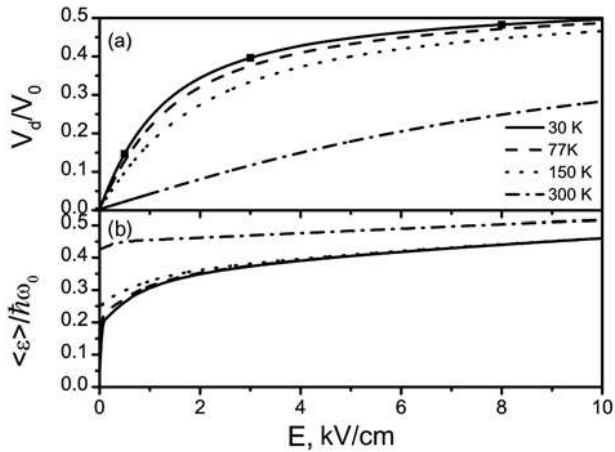}
\vskip-3mm\caption{Dependences of ({\it a})~drift velocity and ({\it
b})~average energy of electrons on the electric field amplitude.
$V_{0}=4\times10^{7}~\mathrm{cm/s} $ is the characteristic velocity.
The other parameters of the material are the same as in Fig.~3.
Points correspond to the field values, at which the distribution
functions depicted in Fig.~3 were calculated  }
\end{figure}

The calculated distributions of electrons in the momentum space along and
across the applied electric field are shown in Fig.~3 (panels $a$ and $b$,
respectively). Already in a field of 0.5~kV/cm, the system of electrons
becomes nonequilibrium, and the difference between the electron
distributions along and across the electric field is well distinguished. The
distribution of electrons along the electric field has an appreciable
asymmetry, which is associated with the existence of two groups of
electrons, namely, low-energy electrons, which are isotropically distributed
in the space of their momenta, and high-energy ones with the momenta
directed along the electric field. As the electric field grows,
the number of such monodirected electrons increases, and the electron
distribution becomes more and more anisotropic. In the perfect streaming
limit, all electrons have their momenta directed along the field, and the
distribution function of electrons along the field has a step-like shape.
From Fig.~3,$a$, it is evident that, for the fields of 3--8~kV/cm, the shape
of the electron distribution function along the field is close to the step-like one.
At the same time, the distribution function of electrons across the field,
$f(p_{x})$, remains symmetric. In the directions perpendicular to the field,
the dominating fraction of electrons have momenta $p_{x},p_{y}<p_{0}$. For
the growing field, one can observe a weak compression of tails of the transverse
component of the distribution function $f(p_{x})$ (see Fig.~3,$b$).

In the fields higher than 10~kV/cm, the streaming effect becomes destroyed,
because electrons penetrate deeper into the active energy range.

\subsection{Steady-state electric characteristics}

The streaming regime can manifest itself through the emergence of
characteristic features in the dependences of stationary electric
parameters on the applied field. Among those, there are weak
dependences of the drift velocity $V_{d}$ and the average energy
$\langle \epsilon \rangle $ on the applied electric field strength
$E$. In Fig.~4, one can easily see that, in the range of low, i.e.
pre-streaming, electric fields 0.1--1~kV/cm, the dependence of the
drift velocity on the field is approximately linear, but becomes
sublinear if the field continues to grow. In the fields 3--10~kV/cm,
the streaming type of electron transport is formed: the drift
velocity practically does not change as the field grows, being close
to $V_{0}/2$, which corresponds to the drift velocity in the perfect
streaming model. A similar scenario is observed for the field
dependence of the average energy. In the developed streaming regime,
the value of $\langle \epsilon \rangle (E)$ approaches that of
$\hbar \omega _{0}/3$, which is characteristic of the perfect
streaming limit. It is worth noting that the dependences $V_{d}(E)$
and $\langle \epsilon \rangle (E)$ qualitatively do not differ
strongly from each other within the temperature interval
\mbox{30--150~K.}

At room temperature, the situation is different. The streaming
regime does not emerge owing to the presence of the strongly
inelastic scattering in the passive energy range, the mechanism of
which consists in the absorption of a polar optical phonon. In the
fields 3--10~kV/cm, the dependence of the drift velocity on the field
remains almost linear, the average energy almost does not change and
remains close to the equilibrium value $3/2\times k_{\rm b}T$.

The emergence of the streaming regime can also be clearly traced by detecting a
nonmonotonous field dependence of the transverse component of the average
electron energy, $\langle \epsilon _{\perp }\rangle =\langle
(p_{x}^{2}+p_{y}^{2})/2m^{\ast }\rangle $, where $p_{x}$ and $p_{y}$ are the
electron momenta in the directions perpendicular to the field. In Fig.~5,
the field dependences of $\langle \epsilon _{\perp }\rangle $ calculated for
various lattice temperatures are depicted. At cryogenic temperatures and in
low fields (less than 1~kV/cm), the value of $\langle \epsilon _{\perp
}\rangle $ grows owing to the heating of the electron gas, which mainly manifests
itself in the isotropic broadening of the electron distribution function in the
momentum space. If the field amplitude increases further, $\langle \epsilon
_{\perp }\rangle $ decreases and saturates in fields that correspond to the
developed streaming, which is associated with the narrowing of high-energy
tails of the electron distribution function in the directions perpendicular
to the field (see Fig.~3,$b$). At room temperature, both $\langle \epsilon
_{\perp }\rangle $ and the total average energy $\langle \epsilon \rangle $
practically do not change as the field grows.

Another attribute of the fact that the electron transport is in the
streaming regime consists in a specific behavior of the field
dependence of the electron diffusion coefficient in the coordinate
space. As was shown above, the electron distribution function across
the field gets narrowed at the streaming formation. As a result, the
diffusion motion of electrons across the field must be inhibited,
which would lead to a reduction of the diffusion coefficient
$D_{\perp }$ in the direction perpendicular to the field. This
feature was discussed in work~\cite{Korotyeyev1}, when analyzing the
streaming effect for the two-dimensional electron gas in the
framework of the Baraff approximation. For today, a lot of
researches have been carried out concerning the diffusion
coefficient in bulk nitride samples~\cite{Varani4} in strong fields
(up to 0.5~MV/cm). At the same time, the interval of moderate
electric fields, at which the streaming effect becomes possible, was
not given a proper attention.

In Fig.~6, the field dependences $D_{\perp}(E)$ calculated for
various lattice temperatures by the Monte Carlo method are shown. In
heating fields of 0--0.25~kV/cm and at cryogenic lattice
temperatures, the magnitude of $D_{\perp}$ drastically grows. If the
field amplitude grows further and the streaming-like distribution of
electrons is formed, the value of $D_{\perp}$ decreases. Starting
from a field of 3~kV/cm, i.e. in the developed streaming regime,
$D_{\perp}$ saturates and approached values of
20--25$~\mathrm{cm}^{\mathrm{2}}/\mathrm{s}$, which are close to or
even lower than the values for the equilibrium diffusion
coefficients---the corresponding values are 13, 57, and
120$~\mathrm{cm}^{\mathrm{2}}/\mathrm{s}$ at lattice temperatures of
30, 77, and 150~K, respectively. At room temperature, $D_{\perp}$ is
almost independent of the field amplitude, varying from
50$~\mathrm{cm}^{\mathrm{2}}/\mathrm{s}$ in the zero field to
30$~\mathrm{cm}^{\mathrm{2}}/\mathrm{s}$ in a field of
10~kV/cm.

Thus, a number of features in the field dependences of electric parameters,
which point to the emergence of the streaming effect, are observed for the
electron gas in a compensated bulk GaN with the concentration of ionized
impurities $N_{i}=10^{16}$~\textrm{cm}$^{-3}$, the electron concentration
$n_{e}=10^{15}$~\textrm{cm}$^{-3}$, in the fields of a few kV/cm, and at
lattice temperatures of 30--150~K. It is worth noting that the
streaming regime can be identified according to the results of experimental
measurements of the drift velocity and the diffusion coefficient of
electrons in the direction perpendicular to the field.

\begin{figure}
\includegraphics[width=\column]{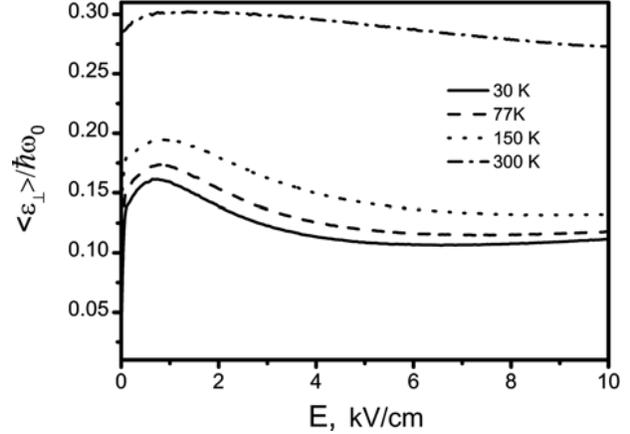}
\vskip-3mm\caption{Transverse component of the average electron
energy as a function of the field  }\vskip3mm
\end{figure}

\begin{figure}
\includegraphics[width=\column]{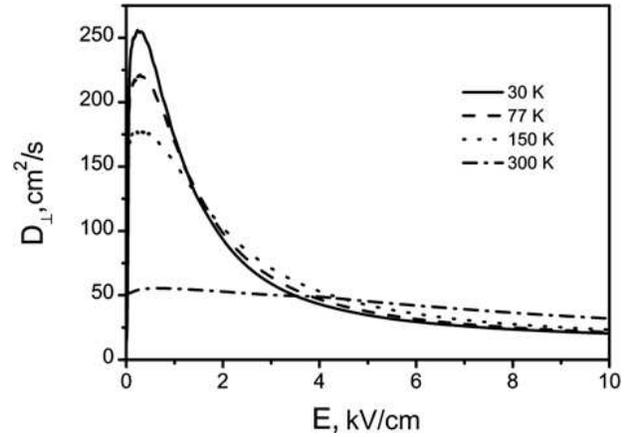}
\vskip-3mm\caption{Diffusion coefficient in the direction
perpendicular to the field as a function of this field. The other
parameters of the material are the same as in Fig.~3  }
\end{figure}
\begin{figure}
\includegraphics[width=\column]{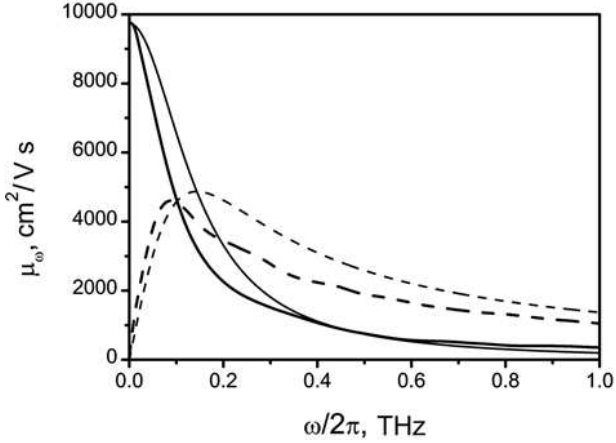}
\vskip-3mm\caption{Dependence of DDM on the frequency calculated in
the framework of the Drude--Lorentz model (thin curves) and by the
Monte~Carlo method (bold curves). Solid curves correspond to
$\mathrm{Re}[\mu_{\omega}]$ and dashed ones to
$\mathrm{Im}[\mu_{\omega}]$. The dc field amplitude $E=0.5$~kV/cm,
$T=30$~K, $N_{i}=10^{16}$~\textrm{cm}$^{-3}$, and
$n_{e}=10^{15}~\mathrm{cm}^{-3}$  }\vskip3mm
\end{figure}

\begin{figure}
\includegraphics[width=\column]{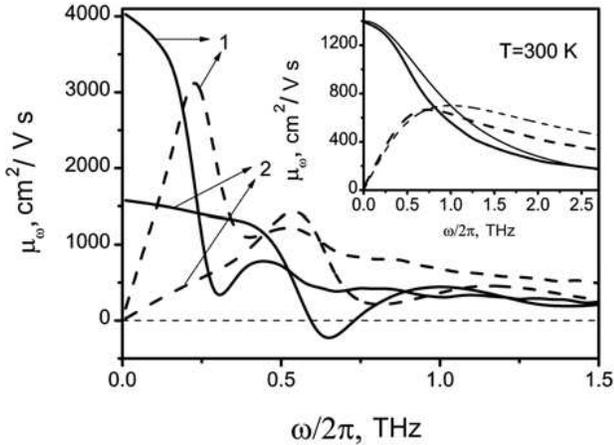}
\vskip-3mm\caption{Dependences $\mathrm{Re}[\mu_{\omega}]$ (solid
curves) and $\mathrm{Im}[\mu_{\omega}]$ (dashed curves) at $T=30$~K
and $E=1.5$ ({\it 1}) and 3~kV/cm ({\it 2}). The spectra
$\mathrm{Re}[\mu_{\omega}]$ and $\mathrm{Im}[\mu_{\omega}]$ at
$E=3$~kV/cm and $T=300$~K are shown in the inset  }
\end{figure}

\section{High-Frequency Electric Characteristics}

\label{secIV}Surely, the most remarkable manifestation of the
streaming phenomenon is the transit-time resonance effect, which is
connected with the emergence of the negative dynamic differential
conductivity as a response to an external high-frequency signal. In
this section, we will discuss the spectra of the dynamic
differential mobility (DDM), $\mu _{\omega }$, calculated by the
Monte Carlo method using the single-particle
algorithm~\cite{Reggiani, Zimerman1, Lebwohl}. We will analyze
comprehensively the conditions of existence for the negative dynamic
differential mobility (DNDM) in a compensated GaN at various lattice
temperatures and various relative orientations of dc and ac electric
fields. The analysis of $\mu _{\omega }$ is started from the case
where a dc field $E$ and a low sinusoidal field with amplitude
$E_{\omega }\ll E$, which are parallel to each other, are
applied to the sample.

\subsection{Dynamic differential mobility}

In Fig.~7, the bold curves correspond to the frequency dependences of $\mu
_{\omega}$ calculated for the low stationary electric field $E=0.5$~kV/cm
using the Monte Carlo method. For comparison, the same dependences
calculated in the framework of the Drude--Lorentz model,
\begin{equation*}
\mu_{\omega}=\mu_{0,E}/(1-i\omega m^{\ast}\mu_{0,E}/e),
\end{equation*}
where $\mu_{0,E}=dV_{d}(E)/dE$ is the zero-frequency differential mobility,
are also shown by thin curves. In the field $E=0.5$~kV/cm, the electron
distribution function still remains quasiisotropic, and, as is seen from
Fig.~7, the shape of the $\mu_{\omega}$-spectrum is close to that of
the Drude--Lorentz one.

If the dc field amplitude increases, the streaming regime starts to
be formed, which manifests itself in the oscillatory dependence of
the real and imaginary parts of DDM on the frequency. In Fig.~8, the
$\mu _{\omega }$-spectra calculated for $E=1.5$ and 3~kV/cm are
depicted. Already at $E=1.5$~kV/cm, the real part of DDM has a
series of minima at the transit-time resonance frequency and its
higher harmonics, but still remains positive. At $E=3$~kV/cm, the
real part of DDM changes its sign and becomes negative in a vicinity
of the fundamental frequency of the transit-time resonance $\omega
_{\rm tr}/2\pi \approx 0.6~$THz. Note that, at room temperature and
in the same electric field, the real and imaginary parts of DDM do
not reveal any features, being well described by the Drude--Lorentz
model. In the inset of Fig.~8, bold (calculation by the Monte Carlo
method) and thin (the Drude--Lorentz model) curves are close to each
other.

Hence, the amplitude and frequency windows of DNDM directly depend
on the applied field magnitude and the temperature of a sample. In
the frequency range, where $\mathrm{Re}[\mu _{\omega }]<0$, the
amplification of a high-frequency signal occurs, which is
proportional to the DNDM amplitude. Therefore, it is expedient to
determine the intervals for temperatures and dc electric fields, at
which the DNDM takes place and reaches minimum negative values.

\subsection{Dependences of the OPTTR effect on the temperature and the field
amplitude}

\label{sec4.2}In Fig.~9, the frequency dependences of
$\mathrm{Re}[\mu _{\omega }]$ are depicted only for the first actual
resonance minimum. At higher-order minima, which are multiples of
the transit-time frequency, the DNDM does not emerge. The most
pronounced OPTTR effect is observed at a temperature of 30~K. The
DNDM appears at $E\approx 2$~kV/cm and reaches a minimum value of
$-250~\mathrm{cm}^{\mathrm{2}}/\mathrm{V/s}$ in a vicinity of the
frequency $\omega /2\pi \approx 0.64~$THz at $E\approx 3$~kV/cm. As
the dc field amplitude increases, the frequency windows of the DNDM
shift toward the high-frequency range and are broaden out. At the
same time, however, the DNDM amplitude decreases. In Fig.~9, it is
clearly seen that, at $T=30$~K, the DNDM appears in the field range
within the limits 2--10~kV/cm and in the frequency interval from
0.38 to 2.5~THz (this interval is confined by a dash-dotted curve).
However, if the field amplitude exceeds 10~kV/cm, a sufficient
number of electrons can penetrate into the active range to violate
the coherent motion of the majority of electrons, so that the DDM
values become positive.

At $T=77$~K, the frequency windows with the DNDM still exist.
However, the DNDM manifests itself much weaker than at $T=30$~K. The
frequency interval, in which the DDM becomes negative, is much
narrower at $T=77$~K than at $T=30 $~K. At the fields within the
interval 2.5--9~kV/cm, the DNDM emerges at frequencies between
approximately 0.56 and 2.2~THz. The largest DNDM amplitude is
realized at $E=4$~kV/cm at a frequency of 0.9~THz and with the
minimum $\mathrm{Re}[\mu _{\omega }]\approx
-100~\mathrm{cm}^{\mathrm{2}}/\mathrm{V/s}$.

At $T=150$~K, the DDM also demonstrates the oscillatory dependence
on the frequency; however, it does not become negative. It is of
interest that, at this temperature, the low-field mobility $\mu
_{0}=9500~\mathrm{cm}^{\mathrm{2}}/\mathrm{V/s}$ (see Fig.~2), which
is almost twice as large as that at $T=30~$K ($\mu
_{0}=5000~\mathrm{cm}^{\mathrm{2}}/\mathrm{V/s}$). Hence, the
conditions for the streaming effect to take place seem to be better
at $T=150 $~K rather than at $T=30$~K, and, consequently, the OPTTR
effect should have manifested itself more strongly just in the
former case. However, the opposite situation is actually observed.
Such a disagreement can be explained by the sensitivity of the OPTTR
effect to the initial broadening of the equilibrium electron
distribution function. For the distribution function with an
anisotropy, which would be sufficient for the appearance of the DNDM
at $T=150$~K, to be formed, a field stronger than that required at
30--77~K has to be applied to the semiconductor. However, in
stronger fields, the DNDM does not appear, because the penetration
of electrons into the active energy range becomes substantial.

\subsection{Anisotropy of dynamic differential mobility}

In Section 4.2, the dependences of the real part of DDM on the
frequency of the ac electric field and the amplitude of the dc one, which are
parallel to each other, were discussed in detail. Under the
streaming conditions, when the distribution function of electrons is
anisotropic, one may expect that the electron response should depend
on the relative orientation of the dc field and the varying signal.

\begin{figure}
\includegraphics[width=\column]{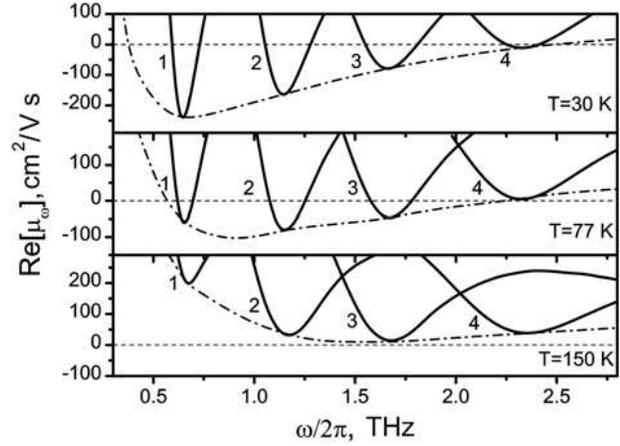}
\vskip-3mm\caption{ $\mathrm{Re}[\mu_{\omega}]$-spectra in a
vicinity of OPTTR frequencies at $E=3$ ({\it 1}), 5 ({\it 2}), 7
({\it 3}), and 9~kV/cm ({\it 4}). The dash-dotted curve is the
envelope of $\mathrm{Re}[\mu_{\omega}]$-minima.
$N_{i}=10^{16}$~\textrm{cm}$^{-3}$ and
$n_{e}=10^{15}~\mathrm{cm}^{-3}$ }\vskip3mm
\end{figure}

\begin{figure}
\includegraphics[width=\column]{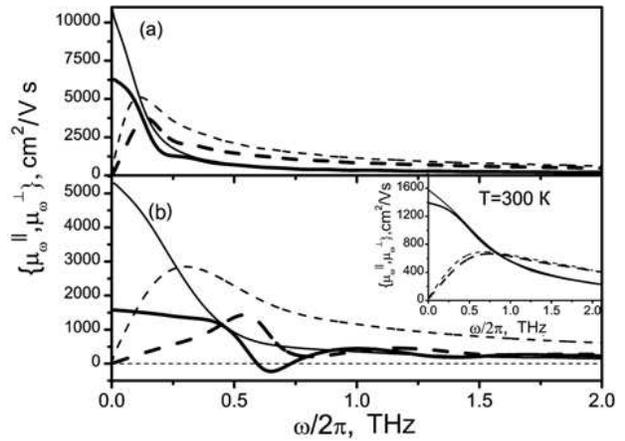}
\vskip-3mm\caption{Spectra of $\mu_{\omega}^{\parallel}$ (bold
curves) and $\mu_{\omega}^{\perp}$ (thin curves) at
$T=30~\mathrm{K}$ and $E=1$ ({\it a}) and 3~kV/cm ({\it b}). The
real and imaginary parts of DDM are shown by solid and dashed,
respectively, curves. The same spectra but at $T=300~\mathrm{K}$ and
$E=3$~kV/cm are shown in the inset  }
\end{figure}

\begin{figure}
\includegraphics[width=\column]{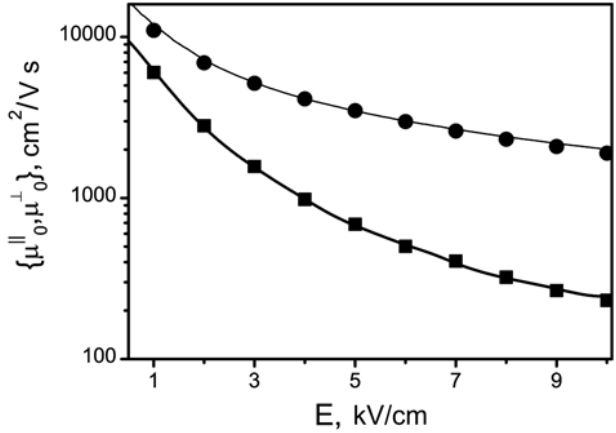}
\vskip-3mm\caption{Dependences $\mu_{\omega}^{\parallel}$ (bold
curves) and $\mu_{\omega }^{\perp}$ (thin curves) calculated from
stationary characteristics $V_{d}(E) $,
$\langle\epsilon_{\perp}\rangle(E)$, and $D_{\perp}(E)$.
$\mu_{0}^{\parallel}$ (squares) and $\mu_{0}^{\perp}$ (circles)
values were calculated by the Monte Carlo method. The lattice
temperature is 30~K  }
\end{figure}

In Fig.~10, the dependences of DDM on the frequency of ac signals
polarized in parallel and perpendicularly to the direction of the dc
electric field are exhibited. The quantities $\mu _{\omega
}^{\parallel}$ and $\mu _{\omega }^{\perp }$ describe the response
of the electron system to the ac signal with the parallel and
perpendicular, respectively, polarizations. At low frequencies, the
$\mu _{\omega }^{\parallel}$- and $\mu _{\omega }^{\perp }$-values
substantially differ from each other both in low fields
($<1~\mathrm{kV/cm}$), in which the distribution function of
electrons is characterized by a moderate anisotropy, and in the
developed streaming regime ($E=3$~kV/cm).

The difference between $\mu _{\omega }^{\parallel}$ and $\mu
_{\omega }^{\perp }$ survives even at the zero frequency, which is
associated with the anisotropy of the electron distribution in the
momentum space. In Fig.~11, the dependences of parallel, $\mu
_{0}^{\parallel}$, and perpendicular, $\mu _{0}^{\perp }$,
components of the differential mobility on the dc field strength $E$ are
shown. These dependences were obtained with the use of the Monte Carlo
method, in which the $\mu _{0}^{\parallel}$- and $\mu _{0}^{\perp
}$-values were calculated for an extremely low frequency. There is
another approximate method for the determination of those
quantities, which uses the values of stationary parameters. In a
given dc field $E$, $\mu _{0}^{\parallel}\approx dV_{d}(E)/dE$. This
relation is exact in the limit of low-signal response ($E_{\omega
}\ll E $). To calculate $\mu _{0}^{\perp }$, the generalized
Einstein relation can be used,
\begin{equation} \frac{D_{\perp }}{\mu _{0}^{\perp }}\approx
\frac{\langle \epsilon _{\perp }\rangle }{e},  \label{Einst}
\end{equation}%
with regard for the dependences $\langle \epsilon _{\perp
}\rangle (E)$ (see Fig.~5) and $D_{\perp }(E)$ (see Fig.~6). The
values calculated for $\mu _{0}^{\parallel}$ and $\mu _{0}^{\perp }$
by the Monte Carlo method coincide with the corresponding values
obtained from the stationary characteristics $V_{d}(E)$, $\langle
\epsilon _{\perp }\rangle (E)$, $D_{\perp }(E)$, and formula
(\ref{Einst}) (see Fig.~11). The fact that both determination
methods bring about identical values for $\mu _{0}^{\perp }(E)$ can
be explained by an almost Maxwellian distribution of electrons in
the direction perpendicular to the dc field.

From Fig.~11, one can easily see that the dependence $\mu
_{0}^{\parallel}(E)$ decreases more rapidly than the $\mu
_{0}^{\perp }(E)$ one with the growth of the dc field amplitude. Such a
behavior of $\mu _{0}^{\parallel}(E)$ can be explained by the fact
that the sublinearity manifests itself much more strongly in the
dependence $V_{d}(E),$ rather than in the $D_{\perp }(E)/\langle
\epsilon _{\perp }\rangle (E)$ one. In fields of 3--8~kV/cm, which
correspond to the developed streaming regime, a considerable
difference between the $\mu _{0}^{\parallel}(E)$- and $\mu
_{0}^{\perp }(E)$-values is retained. Hence, the experimentally
observed substantial anisotropy of the differential mobility in strong
electric fields can serve as an additional proof of the streaming
formation.

In fields of 1--3~kV/cm, the distribution function of electrons in
the momentum space remains symmetric in the direction perpendicular
to the field. Consequently, the frequency dependence of $\mu
_{\omega }^{\perp }$ is well described by the Drude--Lorentz model
and does not reveal an oscillatory behavior typical of $\mu _{\omega
}^{\parallel}$ (see Fig.~10). It is worth to note that there is no
substantial difference between $\mu _{\omega }^{\perp }$- and $\mu
_{\omega }^{\parallel}$-values in a dc field of 1~kV/cm, except for
at very low frequencies. Under the streaming conditions (e.g., at
3~kV/cm), the response of the electron system in the parallel
configuration of the fields $E$ and $E_{\omega }$ differs cardinally
from that in the perpendicular configuration. In the frequency
interval, where $\mu _{\omega }^{\parallel}$ becomes negative and
$\mu _{\omega }^{\perp }$ remains positive, an ac signal with the
polarization perpendicular to the dc field, instead of being
amplified, is effectively absorbed. This effect can be observed in
optical experiments on the transmission of ultrahigh-frequency
(terahertz) radiation through semiconductor structures. If a strong
enough dc electric field is applied to a sample, the latter is
characterized by the anisotropy of DDM and acts as a polarizer for a
non-polarized beam. The efficiency of such a polarizer depends on
the magnitude of applied dc electric field and the sample
temperature. For instance, at $T=300$~K, when the electron
distribution function is still isotropic and the $\mu _{\omega
}^{\perp }$- and $\mu _{\omega }^{\parallel}$-values practically
coincide, the sample does not function as a polarizer.

Recently, the collaborators of the the Terahertz laboratory at the
Montpellier University carried out an experiment, in which they
tried to register the OPTTR effect and find the DNDM by detecting
the amplification of terahertz radiation transmitted through a
heterostructure with GaN~\cite{Varani5}. Unfortunately, reliable
confirmations for the OPTTR effect were not obtained. To elucidate
what one could expect of such experiments, we developed a theory
describing the light transmission through a sample with an active
epitaxial layer of compensated GaN in the OPTTR regime.

\section{\vspace*{-0.5mm}Transmission of High-Frequency Radiation through a GaN
Structure}

\label{secV}In modern experiments, the complicated multilayered structures grown up
on dielectric substrates are used. As a rule, the thickness of the active zone,
a thin GaN layer, has an order of a few micrometers, which is much less than
the wavelength $\lambda _{0}$ of terahertz electromagnetic radiation in
vacuum. In similar structures, the thickness of dielectric substrate is much
thicker than that of GaN layer, being, as a rule, of the same order of
magnitude as $\lambda _{0}$. In this section, we expound the theory of
high-frequency radiation transmission through such structures. While
studying the spectra of the transmission, reflection, and absorption
coefficients for high-frequency radiation, the frequency dependences of DDM
were applied, which were calculated with the use of the Monte Carlo method
(see Section~\ref{secIV}).

\subsection{\vspace*{-0.5mm}Theory of high-frequency radiation transmission through GaN}

Let high-frequency radiation pass through a structure consisting of a
dielectric substrate with the thickness $d_{s}$ and the dielectric constant
$\kappa _{s}$, covered with a delta-like GaN layer. The GaN layer is
characterized by the two-dimensional complex conductivity $\sigma _{\omega
}^{\ast }=\sigma _{\omega }^{\prime }d+id(\sigma _{\omega }^{\prime \prime
}-\kappa _{0}\omega /4\pi )$, where $\sigma _{\omega }^{\prime
}=en_{e}\mathrm{Re}[\mu _{\omega }]$, $\sigma _{\omega }^{\prime \prime
}=en_{e}\mathrm{Im}[\mu _{\omega }]$, and $d$ is the thickness of GaN layer. This
expression takes the displacement current in the GaN layer into account. Let
a plane wave $E_{p,\omega }(y)\exp (-i\omega t)$ with the amplitude
$E_{p,\omega }(y)$ and the frequency $\omega $ fall normally on the
structure surface. The electric field $E_{p,\omega }(y)$ of this wave
satisfies the Maxwell equations,
\[
\frac{d^{2}E_{p,\omega }}{dy^{2}}+\left\{\!\! \begin{array}{cc}
\frac{\omega ^{2}}{c^{2}} & \scriptstyle{y<0} \\
\frac{\kappa _{s}\omega ^{2}}{c^{2}} & \scriptstyle{0<y<d_{s}} \\
\frac{\omega ^{2}}{c^{2}} &
\scriptstyle{y>d_{s}}\end{array}\!\!\right\} E_{p,\omega }=
\]\vspace*{-4mm}
\begin{equation}
=-\frac{4\pi i\omega \sigma _{\omega }^{\ast ,p}}{c^{2}}E_{p,\omega
}\delta (y),  \label{Max}
\end{equation}%
where the subscript $p$ specifies the wave polarization either along
($p=\Vert $) or across ($p=\bot $) the dc electric field. In Eqs.~(\ref{Max}),
the whole structure is supposed to be in vacuum. The solution of system
(\ref{Max}) looks like
\[
E_{p,\omega }(y)=
\]\vspace*{-5mm}
\begin{equation}
=\left\{\!\! \begin{array}{ll} A_{p,\omega }\exp
(ik_{0}y)+B_{p,\omega }\exp (-ik_{0}y), & {y<0},
\\ C_{p,\omega }\exp (ik_{s}y)+D_{p,\omega }\exp (-ik_{s}y), &
{0<y<d_{s}}, \\
F_{p,\omega }\exp (ik_{0}y), & {y>d_{s}},\end{array}\right.
\label{Maxsol}
\end{equation}%
where $k_{0}=\omega /c$ and $k_{s}=\omega \sqrt{\kappa _{s}}/c$ are the
wave numbers of plane waves in vacuum and the substrate, respectively. The
coefficients $A_{p,\omega }$, $B_{p,\omega }$, $C_{p,\omega }$, $D_{p,\omega
}$, and $F_{p,\omega }$ are determined from the following conditions at the
coordinate planes $y=0$ and $y=d_{s}$:
\[
E_{p,\omega}(-0)=E_{p,\omega}(+0), \]\vspace*{-5mm}
\[\frac{dE_{p,\omega}}{dy}(-0)-\frac{dE_{p,\omega}}{dy}(+0)=
\frac{4\pi i\omega\sigma_{\omega}^{*,p}}{c^{2}}E_{p,\omega}(0),
\]\vspace*{-5mm}
\[E_{p,\omega}(d_{s}-0)=E_{p,\omega}(d_{s}+0), \]\vspace*{-5mm}
\begin{equation}
\frac{dE_{p,\omega}}{dy}(d_{s}-0)=\frac{dE_{p,\omega}}{dy}(d_{s}+0).
\end{equation}

After standard transformations, we obtain the following formulas for
the transmission, $T_{p,\omega}$, and reflection, $R_{p,\omega}$,
coefficients:
\[ T_{p,\omega}=\Biggl[\!
\left(\!1+\frac{\Gamma_{p,\omega}^{\prime}}{2}\!\right)^{\!2}+
\frac{\Gamma_{p,\omega}^{\prime\prime 2}}{4}+\]\vspace*{-5mm}
\[
+\frac{(\kappa_{s}-1)(\kappa_{s}-(1+\Gamma_{p,\omega}^{\prime})^{2}-
\Gamma_{p,\omega}^{\prime\prime 2})}{4\kappa_{s}}\times
\]\vspace*{-5mm}
\begin{equation}
\times\sin^{2}(k_{s}d_{s})-\frac{\sqrt{\kappa_{s}}(\kappa_{s}-1)\Gamma_{p,\omega}^
{\prime\prime}}{4\kappa_{s}}\sin(2k_{s}d_{s})\Biggr]^{-1}\!,
\label{Trans}
\end{equation}\vspace*{-5mm}
\[
R_{p,\omega}=\Biggl[\frac{\Gamma_{p,\omega}^{\prime
2}+\Gamma_{p,\omega} ^{\prime\prime 2}}{4}+\]\vspace*{-5mm}
\[
+\frac{(\kappa_{s}-1)(\kappa_{s}-(1-\Gamma_{p,\omega}^{\prime})^{2}-
\Gamma_{p,\omega}^{\prime\prime
2})}{4\kappa_{s}}\sin^{2}(k_{s}d_{s})-\]\vspace*{-5mm}
\begin{equation}
-\frac{\sqrt{\kappa_{s}}(\kappa_{s}-1)\Gamma_{p,\omega}^{\prime\prime}}
{4\kappa_{s}}\sin(2k_{s}d_{s})\Biggr]\times T_{p,\omega}.
\label{Refl}
\end{equation}
The absorption coefficient can be calculated by the formula $L_{p,\omega
}=1-T_{p,\omega}-R_{p,\omega}$ or
\begin{equation}
L_{p,\omega}=\Gamma_{p,\omega}^{\prime}\left(\!
1-\frac{\kappa_{s}-1}{\kappa_{s}}\sin^{2}(k_{s}d_{s})\!\right)
\times T_{p,\omega}.   \label{Loss}
\end{equation}

\begin{figure}
\includegraphics[width=\column]{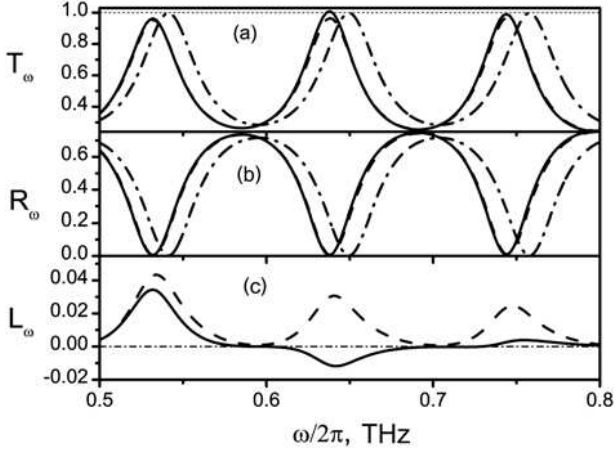}
\vskip-3mm\caption{Spectra of the ({\it a})~transmission, ({\it
b})$~$reflection, and ({\it c})~absorption coefficients for the
parallel (solid curves) and perpendicular (dashed curves) field
configurations. $T_{\mathrm{sub}}$, $R_{\mathrm{sub}},$ and
$L_{\mathrm{sub}}$ spectra are shown by dash-dotted curves. The
parameters of the GaN layer are $N_{i}=10^{16}$~\textrm{cm}$^{-3}$,
$n_{e}=10^{15}$~\textrm{cm}$^{-3}$, $d=10^{-3}$~$\mathrm{cm}$,
$\kappa_{0}=8.9$, and $E=3$~kV/cm. The substrate (sapphire is
supposed) parameters are $d_{s}=0.04$~$\mathrm{cm}$ and
$\kappa_{s}=12$  }\vskip4mm
\end{figure}

\begin{figure}
\includegraphics[width=\column]{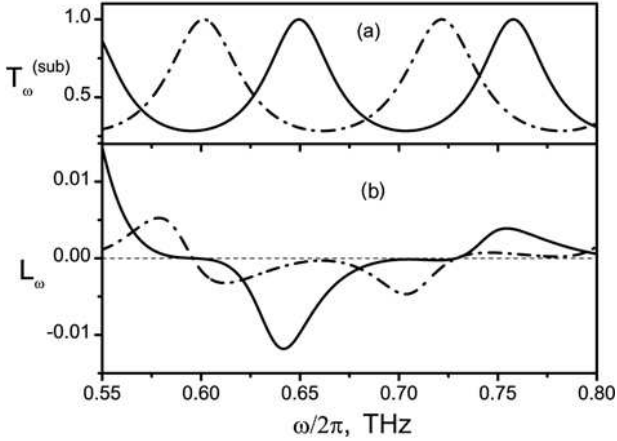}
\vskip-3mm\caption{ ({\it a})~Transmission coefficient spectra for a
single substrate and the ({\it b})~absorption coefficient spectra
for the structure at $d_{s}=0.036$ (dash-dotted curves) and
0.04~$\mathrm{cm}$ (solid curves). The other parameters of the
material are the same as in Fig.~12 }\vskip1mm
\end{figure}

\noindent In all Eqs.~(\ref{Trans})--(\ref{Loss}), the notations
$\Gamma_{p,\omega }^{\prime}$\,\,=\linebreak
$=4\pi\mathrm{Re}[\sigma_{\omega}^{\ast,p}]/c$ and $\Gamma_{p,\omega
}^{\prime\prime}=4\pi\mathrm{Im}[\sigma_{\omega}^{\ast,p}]/c $ are
used. In the absence of a GaN layer, i.e. if there is only the
substrate, the quantities $\Gamma_{p,\omega}^{\prime}$ and
$\Gamma_{p,\omega}^{\prime\prime}$ in formulas
(\ref{Trans})--(\ref{Loss}) equal zero, and the transmission,
reflection, and absorption coefficients are determined in a standard
way, as for a single dielectric wafer, namely,
\[
T_{\rm
sub}=\frac{1}{1+\frac{(\kappa_{s}-1)^{2}}{4\kappa_{s}}\sin^{2}(k_{2}d_{s})},\]%
\[R_{\rm sub}=1-T_{\rm sub},
\]\vspace*{-7mm}
\begin{equation}
 L_{\rm sub}=0. \label{TRLsub}
\end{equation}

\begin{figure}
\includegraphics[width=\column]{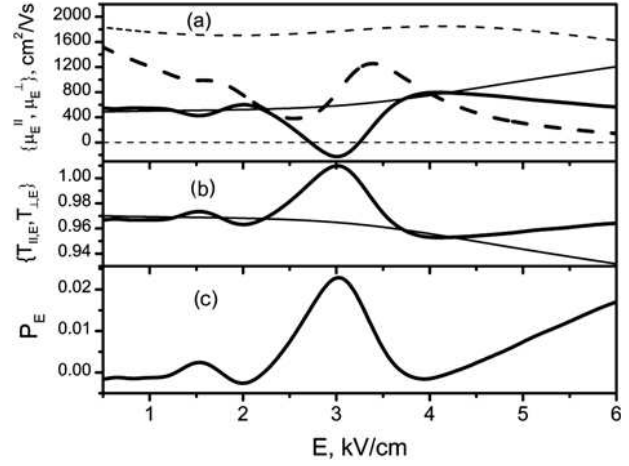}
\vskip-3mm\caption{Dependences of ({\it a})~the real (bold curves)
and imaginary (thin curves) parts of $\mu _{E}^{\parallel}$ (solid
curves) and $\mu _{E}^{\perp }$ (dashed curves), ({\it b}) the
transmission coefficients along ($T_{\parallel,E}$, bold curve) and
across ($T_{\perp ,E}$, thin curve) the applied field, and ({\it c})
the polarization degree $P_{E}$ on the field amplitude at a
frequency of 0.64~THz. The parameters of the substrate and the
active layer are the same as in Fig.~12  }\vskip2mm
\end{figure}

\subsection{Spectra of the transmission, reflection, and absorption coefficients}

The transmission, reflection, and absorption coefficients for
high-frequency radiation (Fig.~12) were calculated in such frequency
intervals and at such values of dc electric field, when the DNDM
manifests itself most strongly (Fig.~10,$b$). If an electromagnetic
wave passes through a single substrate, then, for the given
thickness and the dielectric constant of the latter, we obtain a set
of frequencies $\omega _{r}$ ($r=1,2,3,\mbox{...} $), which
correspond to the Fabry--Perot modes of a plane-parallel dielectric
wafer. Their values are given by the expression $\omega _{r}/2\pi
=cr/(2d_{s}\sqrt{\kappa _{s}})$ and correspond to
$T_{\mathrm{sub}}=1$. The presence of a thin active GaN layer with a
low concentration of electrons on the substrate slightly modifies
the transmission coefficient spectra of the system. The parameters
of the substrate and the active element were so chosen that the DNDM
frequency window should coincide with any of the frequencies $\omega
_{r}$. In the frequency interval 0.6--0.7~THz, the wave with the
polarization along the dc field, when passing through the sample,
becomes amplified. In this frequency interval, the absorption
coefficient becomes negative and reaches a minimum value of $-1.5\%$
at about 0.64~THz (Fig. 12). The negative absorption coefficient
means that the sum of intensities for passed and reflected waves
exceeds the incident wave intensity. Hence, we may say about the
amplification of the electromagnetic field by the active element.
For the sake of comparison, the coefficient of electromagnetic wave
losses at this frequency under the wave reflection from a perfect
silver mirror equals $-0.5\%$. Therefore, the criterion of terahertz
mode excitation in a resonator system composed of two plane-parallel
metallic mirrors can be satisfied despite that the gain factor in
the active layer is~low.

For a wave with the polarization perpendicular to the dc field, the
absorption coefficient is positive, and, hence, there is no
amplification of the electromag\-netic~field.

It should be noted that the coefficient of electromagnetic wave
absorption in the subterahertz frequency range depends on the
substrate parameters. In Fig.~13, the frequency dependences of
$L_{\omega }$ are shown for two samples with identical parameters of
their active elements, but different substrate thicknesses. If the
substrate thickness $d_{s}=0.036$~$\mathrm{cm}$, the frequency
window of DNDM does not coincide with any of the frequencies $\omega
_{r}$ of Fabry--Perot modes. At such non-resonance parameters, the
values of $|L_{\omega }|$ and, hence, the amplification of the
electromagnetic field are substantially lower than those in the
resonance case where $d_{s}=0.04$~$\mathrm{cm}$. Therefore, we may
say about a selective role of the substrate in the amplification of
the electromagnetic field in the subterahertz frequency range.

In experiments dealing with the transmission of radiation through
semiconductor structures, it is much more convenient to measure the
transmission coefficient at a given frequency by varying the
amplitude of the applied dc field. In Fig.~14,$a$, the dependences
of DDM on the applied field amplitude obtained in the parallel, $\mu
_{E}^{\parallel}$, and transverse, $\mu _{E}^{\perp }$, field
configurations at a frequency of 0.64~THz are shown. The real and
imaginary parts of $\mu _{E}^{\parallel}$ have an oscillatory
behavior, and the DNDM is realized in a narrow interval of dc fields
with the amplitudes of about 3~kV/cm. At the same time, the real and
imaginary parts of $\mu _{E}^{\perp }$ almost do not change at that.
At a frequency of 0.64~THz, a substantial difference between $\mu
_{E}^{\parallel}$ and $\mu _{E}^{\perp }$ is observed starting from
the applied field values of \mbox{2--3~kV/cm.}\looseness=1

Experimentally, it is possible to observe the field-induced
difference between the transmission coefficients for electromagnetic
waves with polarizations along, $T_{||,E}$, and across, $T_{\perp
,E}$, the dc field. In Fig.~14,$b$, the bold and thin curves
illustrate the field dependence for $T_{||,E} $ and $T_{\perp ,E},$
respectively. A monochromatic beam, which initially was not
polarized, after having passed through the sample, became partially
polarized. The polarization degree of such a beam, $P_{E}$, depends
on the dc field amplitude, as is shown in Fig.~14,$c$. The quantity
$P_{E}$ is defined as follows: $P_{E}=(T_{||,E}-T_{\perp
,E})/(T_{||,E}+T_{\perp ,E})$. From Fig.~14,$c$, one can see that
the behavior of $P_{E}$ reproduces the oscillatory dependence of DDM
$\mu _{E}^{\parallel}$ on the field, which points to the formation
of the streaming and the appearance of the OPTTR. Beyond the range
of resonance fields, the value of $P_{E}$ monotonously grows, and
the oscillations are absent, which means that the system is not in
the OPTTR. Such a specific dependence of the degree of polarization
of the electromagnetic wave that passed through the sample on the
electric field may also be a characteristic feature of the OPTTR.

\section{Conclusions}

\label{secVI}To summarize, the calculations of the stationary and
high-frequency characteristics of compensated GaN are carried out,
which were aimed at revealing the typical features of the streaming
effect and the conditions needed for the effect to emerge. In
particular, it is found that a strongly anisotropic distribution of
electrons appears in GaN with an impurity concentration of
$10^{16}$~\textrm{cm}$^{-3}$ and an electron concentration of
$10^{15}$~\textrm{cm}$^{-3}$ in the range of applied electric fields
3--8~kV/cm and the temperature interval 30--150~K. Such a
distribution manifests itself as a characteristic saturation in the
dependences of the drift velocity and the total average energy of
electrons on the field. The dependence of the transverse diffusion
coefficient on the field decreases until it reaches the
characteristic saturation. In the framework of the low-signal
response approximation, the spectra of the high-frequency electron
mobility are obtained in the parallel and perpendicular
configurations of the stationary and high-frequency fields. It is
shown that, in the case of the parallel configuration, there exists
a transit-time resonance effect in the frequency range 0.5--2~THz
and the field range 2--10~kV/cm, and the negative dynamic
differential mobility can arise. In the perpendicular configuration,
the negative dynamic differential mobility does not arise, and the
spectrum of the dynamic mobility is close to the
\mbox{Drude--Lorentzian shape.}\looseness=1

On the basis of aforementioned calculations, a theory is developed for the
transmission of terahertz radiation through a structure with an epitaxial
GaN layer. The relative coefficient of terahertz radiation amplification by
the structure operating in the transit-time resonance regime is calculated.
For a single passage of the wave through the GaN structure, the maximum of
the relative amplification coefficient is equal to 1.5\%, which is three times
as large as the losses obtained at the reflection of the same wave from
metallic mirrors. It is shown that the anisotropy of the dynamic mobility leads
to a dependence of the transmission coefficients on the incident wave
polarization. The polarization degree of the wave that passed through the
structure can be controlled by changing the magnitude of applied electric
field.

\vskip3mm The authors are sincerely grateful to Prof. V.O.~Kochelap
(Institute of Semiconductor Physics, National Academy of Sciences of
Ukraine, Kyiv) and Prof. L.~Varani (Montpellier University, France)
for their interest in our researches and their active participation
in the discussion of various aspects of this work. The calculations
were carried out, by using a computer cluster at the Institute of
Semiconductor Physics (Kyiv) in the framework of the State
goal-oriented scientific and engineering program on introducing grid
technologies for 2009--2013.

\vspace{-3mm}
\rezume{Г.I. Сингаївська, В.В. Коротєєв}{%
ЕЛЕКТРИЧНI ТА ВИСОКОЧАСТОТНI ВЛАСТИВОСТI\\ КОМПЕНСОВАНОГО GaN В
УМОВАХ\\ ЕЛЕКТРОННОГО СТРИМIНГУ} {Проведено детальний аналiз умов
iснування стримiнгу і ефекта прольотного резонансу на оптичних
фононах у компенсованому об'ємному GaN. Методом Монте-Карло
проведено розрахунки високочастотної диференцiальної рухливостi.
Показано, що при низьких температурах ґратки 30--77 К в електричних
полях 3--10 кВ/см в терагерцовому дiапазонi частот може iснувати
динамiчна вiд'ємна диференцiальна рухливiсть. Виявленi новi ознаки
ефекту стримiнгу -- анiзотропiя динамiчної диференцiальної
рухливостi i особлива поведiнка коефiцiєнта дифузiї у
перпендикулярному до постiйного електричного поля напрямку.
Побудовано теорiю проходження терагерцового випромiнювання через
структуру з епiтаксiйним шаром GaN. Отримано умови пiдсилення
електромагнiтних хвиль в дiапазонi частот 0,5--2 ТГц. В електричних
полях, бiльших, нiж 1 кВ/см, спостерiгається поляризацiйна
залежнiсть коефiцiєнта проходження випромiнювання через
струк\-туру.}

\end{document}